\documentclass[draft,jgrga]{agutex2015}

\usepackage{graphicx}

 \setkeys{Gin}{draft=false}

\authorrunninghead{OBENBERGER ET AL.}

\titlerunninghead{Radio Meteor Rates and Spectra }

\authoraddr{Corresponding author: K.S. Obenberger,
(obenken@icloud.com)}

\begin{document}

\title{Rates, Flux Densities, and Spectral Indices of Meteor Radio Afterglows}

\authors{K. S. Obenberger\altaffilmark{1}, J.D. Dowell\altaffilmark{2}, P.J. Hancock\altaffilmark{3}, J. M. Holmes\altaffilmark{1}, T.R. Pedersen\altaffilmark{1}, F.K. Schinzel\altaffilmark{2}, G.B. Taylor\altaffilmark{2}}

\altaffiltext{1}{Space Vehicles Directorate, Air Force Research Laboratory, Kirtland AFB, New Mexico, USA}
\altaffiltext{2}{University of New Mexico, Albuquerque, NM, USA}
\altaffiltext{3}{International Centre for Radio Astronomy Research (ICRAR), Curtin University, Bentley, WA, Austrailia}

%
%



\begin{abstract}
Using the narrowband all-sky imager mode of the LWA1 we have now detected 30 transients at 25.6 MHz, 1 at 34 MHz, and 93 at 38.0 MHz. While we have only optically confirmed that 37 of these events are radio afterglows from meteors, evidence suggests that most, if not all, are. Using the beam-forming mode of the LWA1 we have also captured the broadband spectra between 22.0 and 55.0 MHz of four events. We compare the smooth, spectral components of these four events and fit the frequency dependent flux density to a power law, and find that the spectral index is time variable, with the spectrum steepening over time for each meteor afterglow. Using these spectral indices along with the narrow band flux density measurements of the 123 events at 25.6 and 38 MHz, we predict the expected flux densities and rates for meteor afterglows potentially observable by other low frequency radio telescopes. 
\end{abstract}

\begin{article}

\section{Introduction}
Recently, \citet{Obenberger14} discovered that the trails left by bright meteors (fireballs) occasionally emit a radio afterglow in the upper High Frequency (HF; 3 to 30 MHz) and lower Very High Frequency (VHF 30 to 300 MHz) bands.  This self-emission, which has been observed to last up to several minutes, is distinct from the well understood phenomenon of meteor trail reflections. While the radio emission has been observed between 25.6 and 55.0 MHz, continuous broadband measurements have only been made between 37 and 55 MHz. Most events are recorded using the narrowband ($\sim$ 100 kHz) Prototype All-Sky Imager (PASI), a backend correlator/imager of the first station of the Long Wavelength Array (LWA1) \citep{Taylor12,Ellingson13,Obenberger15a}. PASI images have an extremely large field of view of $ 1 \pi$ sr ($\sim$ 10$^{4}$ deg$^{2}$ ) but have limited sensitivity due to the narrow bandwidth, which also prevents any spectral characterization of the events. At the time this paper was written, PASI had detected 30 events at 25.6 MHz, 1 at 34 MHz, and 93 at 38.0 MHz, the majority of these events are thought to be afterglows from meteors. 

Currently the only way to make broadband measurements is with the beamformer mode of the LWA1. This mode can create up to 4 simultaneous beams, each with two tunings of up to 19.6 MHz ($\sim$18 MHz usable), but each beam only has a field of view of $\sim$ 50 deg$^{2}$. This is considerably smaller than the all-sky imager mode, and therefore results in much fewer detections. 

Beamformed measurements of two meteor radio afterglows recorded on October 17, 2014 (M1) and October 26, 2014 (M2) were reported in \citet{Obenberger15b}, and showed that the spectrum is steep, increasing in flux density at lower frequencies. It was also shown that the spectrum of M1 contained dynamic structure in the form of repeating frequency/time dispersed pulses. While the bulk of the emission is largely unpolarized, these sweeps did contain significant linear polarization (Stokes Q). Unfortunately full polarization parameters were not recorded so the amount of Stokes U and V are not known. Also the sweeps were not smooth in spectrum, rather they displayed narrow (several MHz) regions of enhanced emission. On the other hand, M2 only contained smooth spectrum, unpolarized emission, with no sign of the polarized type observed in M1.

The measurements from \citet{Obenberger15b} were made using two tunings, one centered at 45.45 MHz and the other at 65.05 MHz. The upper tunings of each observation were rendered unusable because the beamformer did not have enough dynamic range to properly record the extremely bright forward scatter of analog TV channels. Any channel that did not contain a transmitter was compressed to almost zero value, therefore no useful broadband data were recorded. The lower tuning, however, made relatively clean measurements of the spectrum between 36.5 and 54.0 MHz, with only a few moments of compression. 

The steady increase of flux density at lower frequencies presumably flattens out or turns over somewhere below the 36.5 MHz lower boundary of the M1 and M2 measurements. If the Langmuir wave hypothesis introduced in \citet{Obenberger15b} is correct, then the waves with frequency below the electron/neutral collision rate would be critically damped. At 90 km the collision frequency is $\sim$ 0.5 MHz, and one would expect a sharp cutoff near that frequency. 

At higher frequencies the unpolarized emission for M1 and M2 decayed to the point of undetectability above 52 MHz. Since there is no evidence of a sharp cutoff, it is not unreasonable to assume that the spectra follow a slow approach to zero, perhaps following a power law. Fitting the spectrum would provide numerical parameters that could be used as a test for any future theoretical model. 

Furthermore, a fitted spectrum would allow for a higher frequency extrapolation of the flux density, which would be useful for other facilties, such as the Murchison Wide Field Array (MWA) \citep{Tingay13} in Australia, the Amsterdam-ASTRON Radio Transients Facility and Analysis Center (AARTFAAC), based on the Low Frequency Array (LOFAR) \citep{Prasad14,Haarlem13} in the Netherlands, and two additional Long Wavelength Array stations located at Owens Valley Radio Observatory (LWA-OVRO; www.tauceti.caltech.edu/lwa/array.html) in California and Sevilleta National Wildlife Refuge (LWA-SV) in New Mexico. These telescopes operate at or just above the LWA1 frequency range, have extremely large fields of view, and have comparable or better sensitivity than the LWA1. In addition MWA, AARTFAAC, and LWA-OVRO all have higher angular resolution than the LWA1, creating an excellent opportunity to further the study of meteor radio emission. An extrapolated spectrum along with event rates would give researchers at these facilities an idea of what sensitivities and amount of observing time they might need in order to detect meteor radio afterglows.

Smooth spectrum radio sources are typically fit to a power law, where the spectrum is parameterized with a spectral index $\alpha$. The spectral index is related to the flux density, $S$ by:

\begin{equation}
S \propto \nu^{\alpha},
\end{equation}
 
where $\nu$ is the frequency. At the LWA1 frequency range, a typical spectral index for an astrophysical source is $\sim$ -1, but the spectral index is not necessarily constant over all frequencies. Rather it can be a function of frequency, and is given by :

\begin{equation}
\alpha(\nu) = \frac{\partial \log S(\nu)}{\partial \log \nu}
\end{equation}

\citet{Obenberger15b} did not include a power law fit for the spectra of the two radio afterglows. This analysis was excluded from the paper because the spectrum of M1 and the second meteor (M2) contained several effects that made comparison difficult. As mentioned above, the extremely bright forward scatter of several transmitters put the beamformer into compression, introducing a number of broadband dips in the spectrum. Furthermore, the polarized emission from M1 added spectral structure to several seconds of the event, rendering nearly half of the event impossible to fit. Considering these sources of error, we determined that any results would be difficult to interpret. 

On December 18, 2015 and February 12, 2016 we measured the spectrum of two more meteors afterglows (M3 and M4) at slightly lower frequencies, with significantly cleaner spectra and no beamformer compression. We can compare the smooth spectral components of all four events, and by fitting these spectra to a power law we can predict what flux densities MWA, LOFAR, LWA-OVRO, LWA-SV, and AARTFAAC should expect to see for meteor afterglows at higher frequencies. In addition, we can use the large number of events recorded by PASI to compute rates for events brighter than our worst sensitivity thresholds at 38.0 and 25.6 MHz, and extrapolate these thresholds to predict rates at higher frequencies. These predictions will provide useful values for other observatories to probe the physics of this mysterious phenomenon.

\section{Observations}

As described in \citet{Obenberger15a} PASI, produces near real time images of the sky with all 4 Stokes parameters. PASI can be tuned to any frequency between 10 and 88 MHz and the images cover 75 kHz of bandwidth. The majority of PASI operation has been centered at 38.0 MHz, with 12,876 hours recorded there. Other heavily used frequencies include; 25.6 MHz (1,135 hours), 52.0 MHz (1,779 hours), and 74.0 MHz (1,464 hours). We are now just starting to record at 34 MHz, with only a few hundred hours recorded there.

As standard procedure during PASI operation, we simultaneously form 3 beams pointed around zenith at azimuths of 60$^{\circ}$, 180$^{\circ}$, and 240$^{\circ}$, each with elevations of 87$^{\circ}$. Each beam has two tunings covering 19.6 MHz of bandwidth with 1024, 19.14 kHz channels and has 40 ms of time resolution \citep{Obenberger15b}. Until February of 2016 the beam observations contained limited polarization information, where only Stokes I (total power) and Q (N/S E/W linear polarization) are computed. We recently modified this observations to keep all 4 Stokes parameters to better compare with the data taken with PASI.

When a meteor afteglow, detected with PASI, is observed passing through a beam, a broadband spectrum is extracted. Currently this has occurred four times. The first two (M1 and M2) occurred on October 17 and 26, 2014, while PASI was centered at 38 MHz, the third (M3) occurred on December 18 2015, while PASI was tuned to 25.6 MHz, and the fourth (M4) occurred on February 12, 2016, while PASI was tuned to 34.0 MHz. M1 was captured in two of the beams, providing independent spectral measurements from two different regions of the plasma trail, but M2, M3, and M4 were only recorded in 1 beam. 

While the beam observations for M1 and M2 had their two tunings centered at 45.45 and 65.05 MHz, the M3 and M4 observation was recorded with center frequencies of 32.0 and 46.0 MHz. The meteor search campaign was shifted to lower frequencies based on the observation that the flux density of meteor afterglows increases dramatically at lower frequencies \citep{Obenberger15b}. Therefore, we would get higher signal to noise and presumably increase our detection rate.

M2 occurred just before sunrise and therefore could not have been detected optically. M1, on the other hand occurred at night and was confirmed by the NASA Fireball Network (http://fireballs.ndc.nasa.gov), as reported in \citet{Obenberger15b}. M3 was not detected by the Fireball Network, but was confirmed with an all-sky optical camera we operate near LWA-SV. M4 was neither recorded by the Fireball Network nor the LWA-SV camera, but was confirmed with an all-sky video camera operated by John Briggs, in the town of Magdalena, NM (private communication).

\section{Analysis}

\subsection{Image Analysis}\label{section:ImageAnalysis}

Currently PASI images have recorded 30 transient events at 25.6 MHz, 1 at 34 MHz, and 93 at 38.0 MHz, using the detection process described in \citet{Obenberger14}. Of these 124 events, 76 have occurred at night, and of these, 37 have been confirmed by the NASA Fireball Network cameras in southern NM, Sky Sentinel LLC cameras in northern NM, the camera operated by the authors at LWA-SV, and the camera operated by John Briggs in Magdalena, NM. With a $\sim 50\%$ detection rate, we can estimate that at least half of the observed radio events are indeed meteors. We note that the fireball video cameras/software are only sensitive to bright events with absolute meteor magnitudes brighter than $-2$ \citep{Brown10}, and the optically confirmed events tended to have higher radio flux densities than the unconfirmed events. Therefore, assuming a relation between radio and optical brightness, many of the optically undetected events may be afterglows from dim meteors. This is an open area of research with results to be published soon. 

When comparing light curves, we find 92 of the 124 radio transients display similar light curves characterized by a fast rise and slow decay, in total lasting 20 to 250 seconds (see \citet{Obenberger14} for examples). Moreover, 61 events have been observed to have elongated trail-like structure in the images, suggesting a meteor origin. At least one of these two properties are found among 99 events, 33 of which were optically confirmed events.

The remaining 25 events were point-like and either displayed light curves that did not follow the typical fast rise slow decay pattern or were too short ($\leq 15$ s) to be characterized. Being irregular, it is difficult to guess how many of these are meteor afterglows, although 4 have been confirmed optically. 

\begin{figure}
	\centering
	\includegraphics[width = 5.5in]{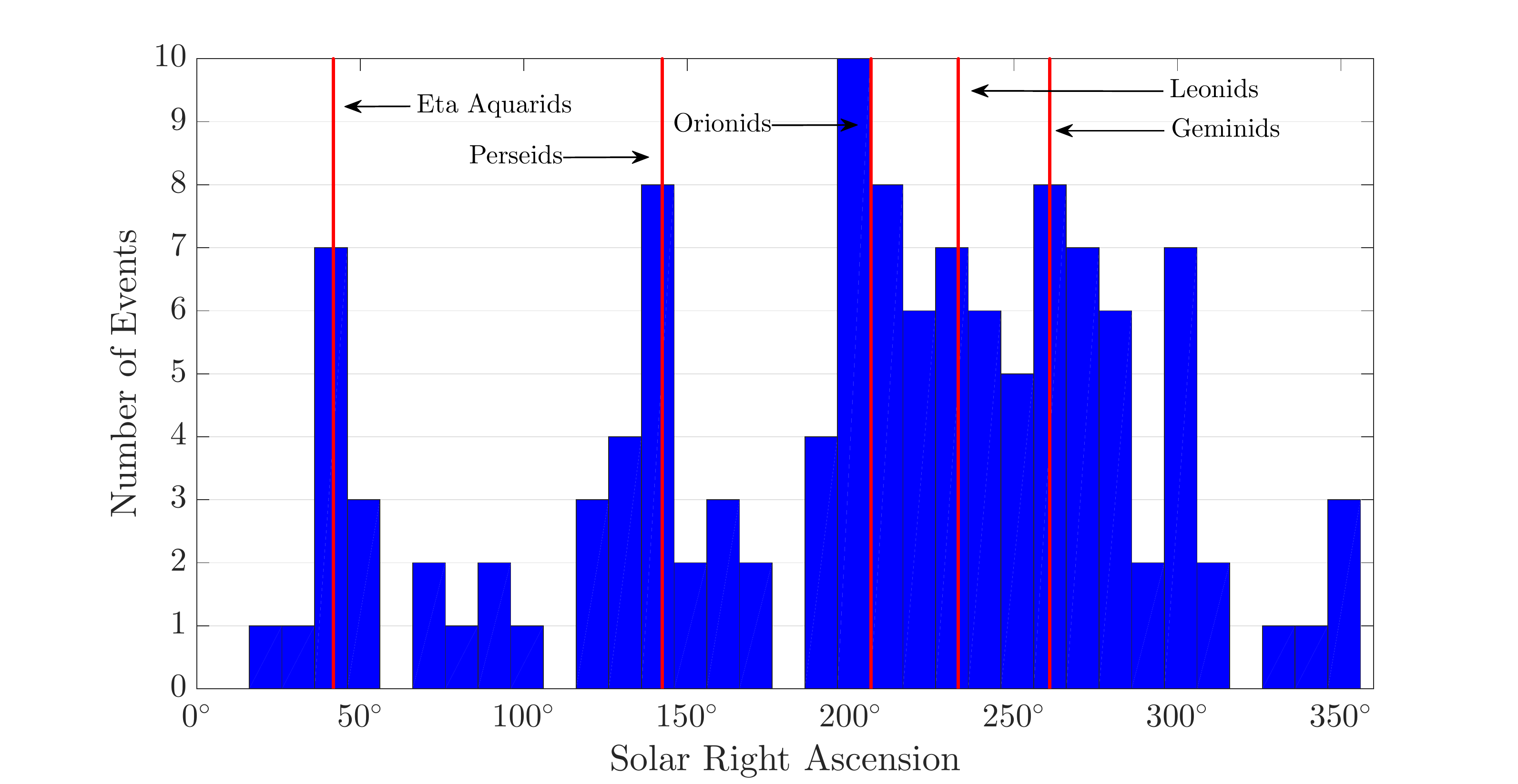}
	\caption{Histogram of transient events as a function of Solar Right Ascension. Several meteor showers have also been plotted to show the correlation between the transients and meteor showers. Correlation is best seen for more isolated showers such as the Eta Aquarids and the Perseids. }
	\label{fig:shower_hist}
\end{figure}

Based on these observations we estimate that $> 85\%$ of the transients are radio afterglows from meteors.  This estimate is further supported by Figure \ref{fig:shower_hist}, which shows strong correlation between the occurrence of the 124 events and the peaks of some major meteor showers. For practical purposes of this paper we will assume that 100\% of the observed radio transients are meteor afterglows.

As mentioned above, of the 124 events, 4 have occurred near zenith during simultaneous beamformed observations. Analysis of the PASI images from M1 and M2 is described in \citet{Obenberger15b}. Similar to the majority of  radio afterglows we observe with PASI, these events displayed a moderately fast rise and slower decay and lasted for several tens of seconds. Also there was no sign of linear or circular polarization. 

The radio detection of M3 only differs from the M1 and M2 in that it was recorded with PASI at 25.6 MHz instead of 38.0 MHz. M4 (recorded at 34 MHz) on the other hand was very short, only being present in two 5 second integrations of PASI. M4 also displayed a significant amount ($\sim$ 50\%) of broadband linear polarization in Stokes U. 

The PASI images of M3 showed no detectable elongation, meaning that compared to the beam size it can be considered a point source. On the other hand, M1, M2, and M4 were observed to have some amount of elongation within the PASI images, which makes the analysis of these spectra slightly more complicated.

\citet{Obenberger15a} describes a method to flux calibrate PASI images. Following this procedure we get flux densities of 730 $\pm 50$ (38.0 MHz), 620 $\pm 50$ (38.0 MHz), 1800 $\pm140$ (25.6 MHz), and 480 $\pm 50$ Jy (34.0 MHz) for the brightest regions of M1, M2, M3, and M4 where 1 Jy = 10$^{-26}$ W m$^{-2}$ Hz$^{-1}$. Uncertainties are derived from the image noise and a 15\% error in the calibrator source Cygnus A \citep{Obenberger15a}.

\begin{figure}
	\centering
	\includegraphics[width = 5.5in]{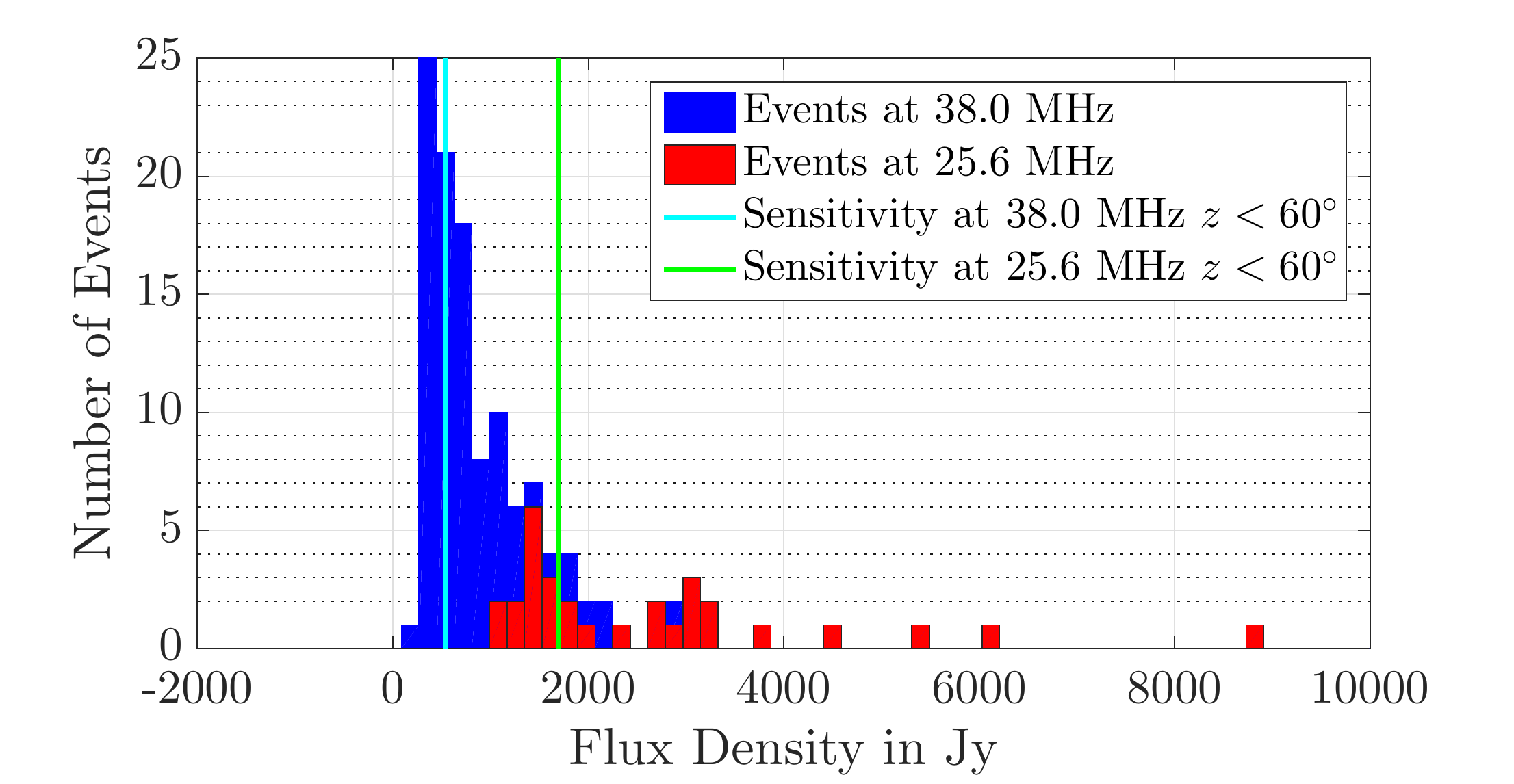}
	\caption{Histogram of events as a function of Flux density for 38 and 25.6 MHz. Events are plotted showing the total number for each frequency. Also plotted are the 6 $\sigma$ image sensitivity for zenith angles $z<60^{\circ}$. All events occurring above these flux densities during observations should have been recorded.  }
	\label{fig:events_hist}
\end{figure}

We have also followed this procedure to extract peak flux densities for all 93 events at 38.0 MHz and 30 events at 25.6 MHz. Figure \ref{fig:events_hist} shows histograms of these events as a function of flux density. Figure \ref{fig:events_hist} also shows the 6 $\sigma$ image sensitivity for zenith angles $z<60^{\circ}$, which are 540 Jy at 38.0 MHz, and 1700 Jy at 25.6 MHz. 

We can take these flux densities as detection thresholds, where all events occurring during PASI observations that were above these flux densities should have been detected. Therefore, we can compute crude rate densities for events brighter than these thresholds. We have detected 60 events above 540 Jy at 38 MHz, where we have observed for a total of 12,876 hours (1.5 years). With $\sim 1 \pi$ sr of sky coverage we can set a rate of $\sim$ 40 events year$^{-1}$ $\pi$ sr$^{-1}$. Similarly for events above 1000 Jy at 38 MHz we can set a rate of $\sim$ 15 events year$^{-1}$ $\pi$ sr$^{-1}$. While the number of events at 25.6 MHz is much lower, we can still compute a rate density for events like M3 of 1800 Jy at 25.6 MHz. With a total of 17 events with flux densities greater than 1800 Jy and 1134.7 hours (0.13 years) of observing time, we get a rate density of $\sim$ 130 events year$^{-1}$ $\pi$ sr$^{-1}$.

These rate density estimates should be understood only as rough estimates for observing, not as an absolute calculation. They were estimated with the assumption that events occur randomly throughout the LWA1 observations. However, we know the rate is not purely random; Figure \ref{fig:shower_hist} shows that events most often occur during meteor showers. Therefore, if a telescope is observing during a meteor shower, the probability of detecting an event is much higher than the rate density would suggest. The exact probability associated with each shower is not known, and most likely changes from year to year. However, if a telescope similar to the LWA1 observes for an entire calendar year, the observed rates should be similar to those reported here.

\subsection{Bandpass Correction of the Spectra}
The bandpass correction for the M1 and M2 spectra was described in \citet{Obenberger15b}. The first step is to remove the frequency dependent response of the antennas and filters, which have been modeled and divided out. Once these instrumental effects are removed we then need to remove the spectrum of the background sky. This is done simply by fitting a smoothing spline to the median spectrum from a few seconds before or after the afterglow. This approach assumes that the spectrum and total power of the background sky in the beam does not significantly change on the timescale of the afterglow. This was effectively true for M1, M2, and M4 but M3, which had a longer duration and occurred closer to the Galactic plane, had a significant amount of background power decrease. To remove this we subtracted the spectrum from the previous day at the same local sidereal time. This method provided a temporally stable removal of the background sky, leaving only a minor spectral residual that was constant over time and easily corrected with a smoothing spline fit from the moments before M3. This method worked well to remove the drifting background sky, but it inherently increased the noise level by $\sqrt{2}$ and still required a smoothing spline fit removal. To maintain high signal to noise, this method was not used on M1, M2, and M4 since the background sky was effectively constant.

The next step is to correct the beam response to the location and shapes of the meteors. \citet{Obenberger15b} described a process where the beam contribution to the spectra was removed for both M1 and M2. This model assumed that the meteor trails were much longer than the width of the beam and uniform in brightness, and used PASI to find the distance away from the beam center. This approach was sufficient for the analysis of that study, but in order to better characterize the spectra, we have reanalyzed those models assuming a length and brightness profile measured from the PASI images, rather than using an infinitely long, uniform brightness trail. While the differences between these two approaches are quite small, the more accurate our model is, the better we can extrapolate the spectra to other frequencies.

\begin{figure}
	\centering
	\includegraphics[width = 5.5in]{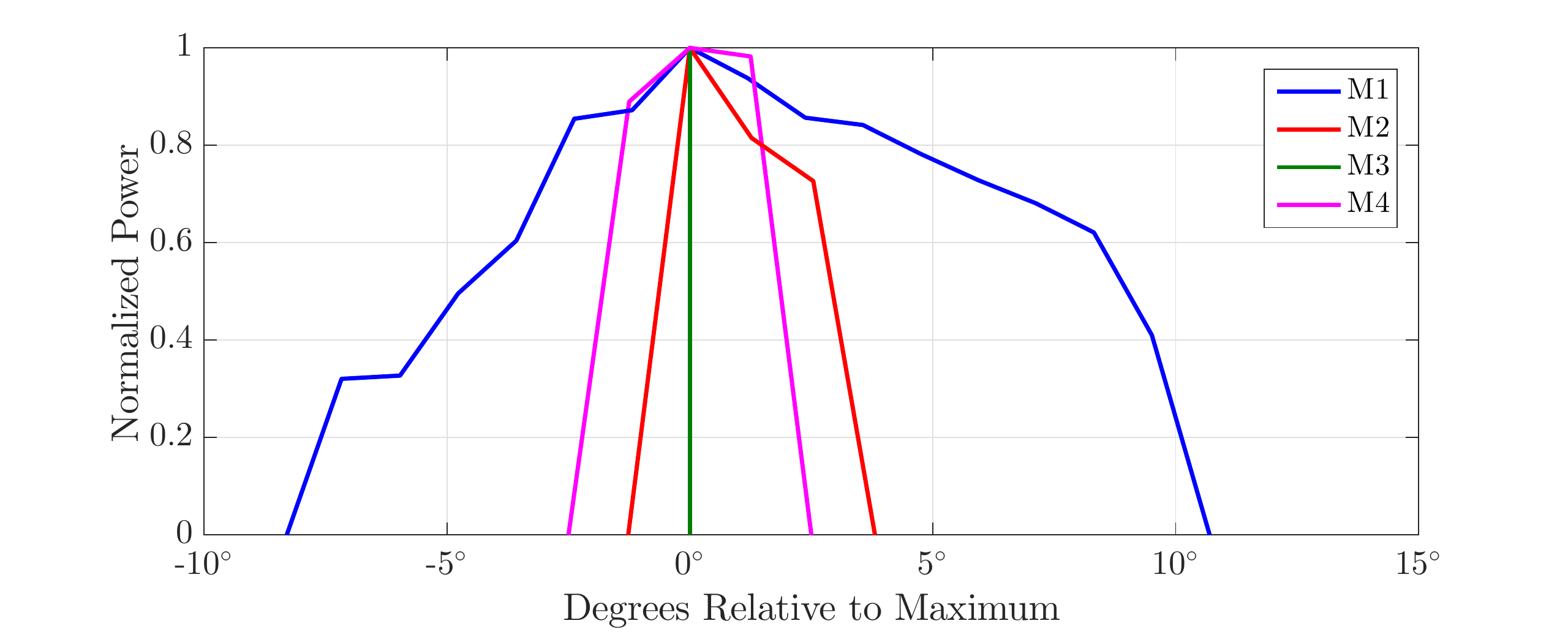}
	\caption{Normalized Brightness profiles along the plasma trails for M1, M2, M3, and M4. Size is measured in degrees relative to the brightest point. M3 is considered point source and is therefore it is represented as a delta function. }
	\label{fig:FB_1_2_profile}
\end{figure}

Figure \ref{fig:FB_1_2_profile} shows the normalized brightness profiles for M1, M2, M4 that were measured with PASI; we assume the profiles for M3 is just point source since it was unresolved in PASI. To find the beam contribution to each spectrum we convolve the measured profiles and locations with the frequency dependent beam shape, which is described by: FWHM$\sim c/ \nu D$, where FWHM is the full width at half maximum, $c$ is the speed of light, $\nu$ is the frequency, and $D$ is the diameter of the telescope. For the LWA1, $D \sim$ 100 m. We compute the convolutions at 13 different frequencies ranging from 20 to 80 MHz. For each meteor we fit the flux response as a function of frequency with a 3rd order polynomial. These fits are then divided out from the spectrum of each meteor.

\subsection{Spectral Indices}

\begin{figure}
	\centering
	\includegraphics[width = 6in]{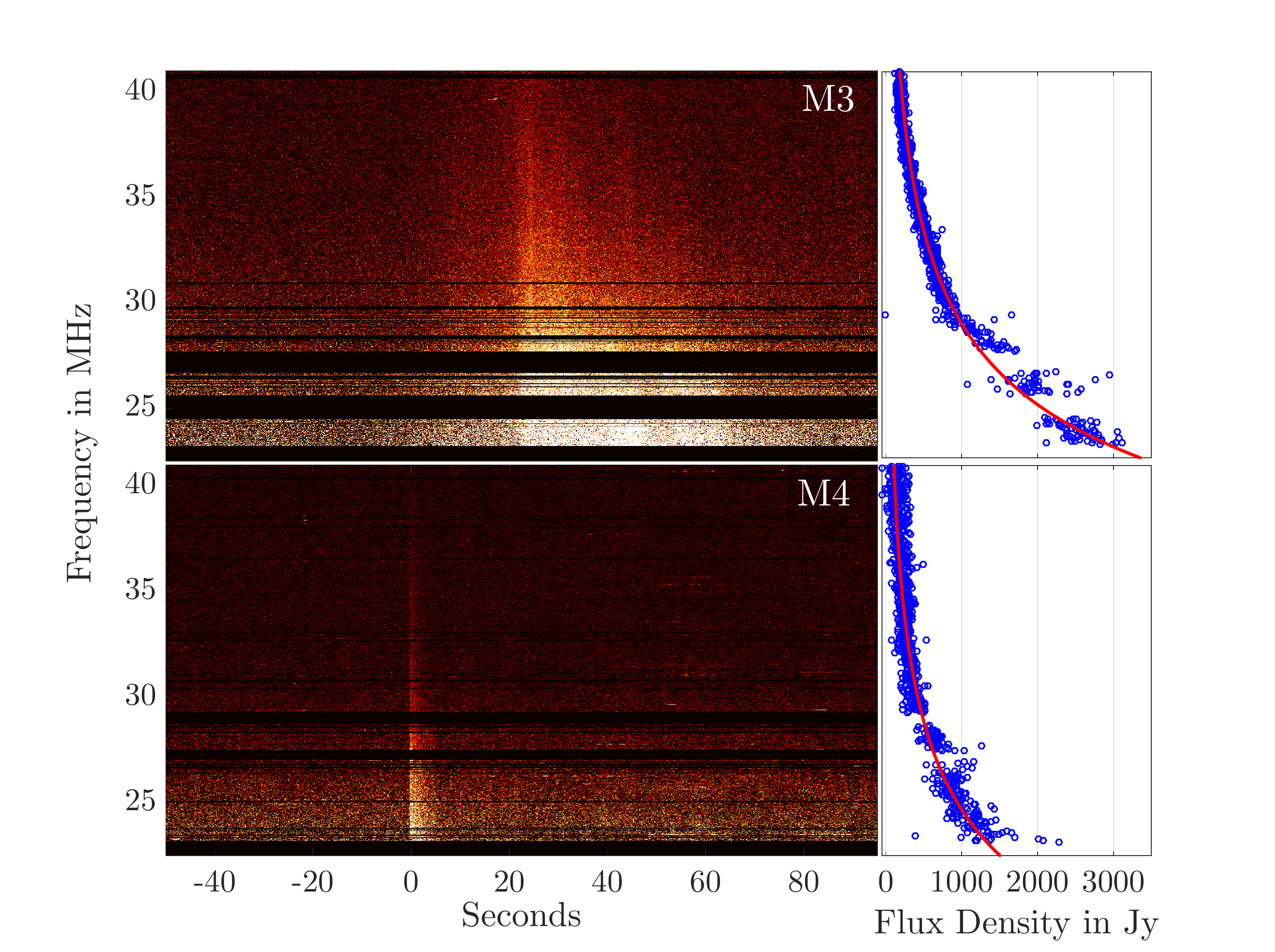}
	\caption{Left: The dynamic spectra of M3 and M4 shown from 22.5 to 41 MHz, shown on the same color scale and the time axes are referenced to the beginning of each radio afterglow. Right: 2 second snapshot spectra taken at the time of peak intensity of M3 and M4. Power law fits are overlaid in red, with a spectral indices of -4.8 and -4.4 for M3 and M4. }
	\label{fig:dSpec_specsnap}
\end{figure}

With the bandpass corrected, we have averaged the 40 ms time resolution dynamic spectra down to 2 second integrations. We have fit each of the 2 second spectrum samples to a power law, and have extracted the spectral indices. Figure \ref{fig:dSpec_specsnap} shows the dynamic spectra for M3 and M4 along with a 2 second example snapshot spectra from the peak intensity overlaid with the power law fits, with spectral indices of -4.8 and -4.4 respectively.

A power law fit M2, M3, and M4 well throughout their duration, with the coefficient of determination, $R^{2}$, having values ranging from 0.8 to 0.96. M1, on the other hand, proved to be much more difficult to fit. Both M1 beams contained a significant amount of polarized emission, which did not have a smooth spectrum. This emission was present throughout the duration of the afterglow until about the 25th second. We still attempted to fit the first 12 seconds seen in each beam, but it was difficult to isolate and exclude the polarized emission contribution. We ignored the 13th through 23rd seconds because very bright polarized emission combined with severe saturation of the beamformer due to bright reflections made fitting impossible.

\begin{figure}
	\centering
	\includegraphics[width = 6in]{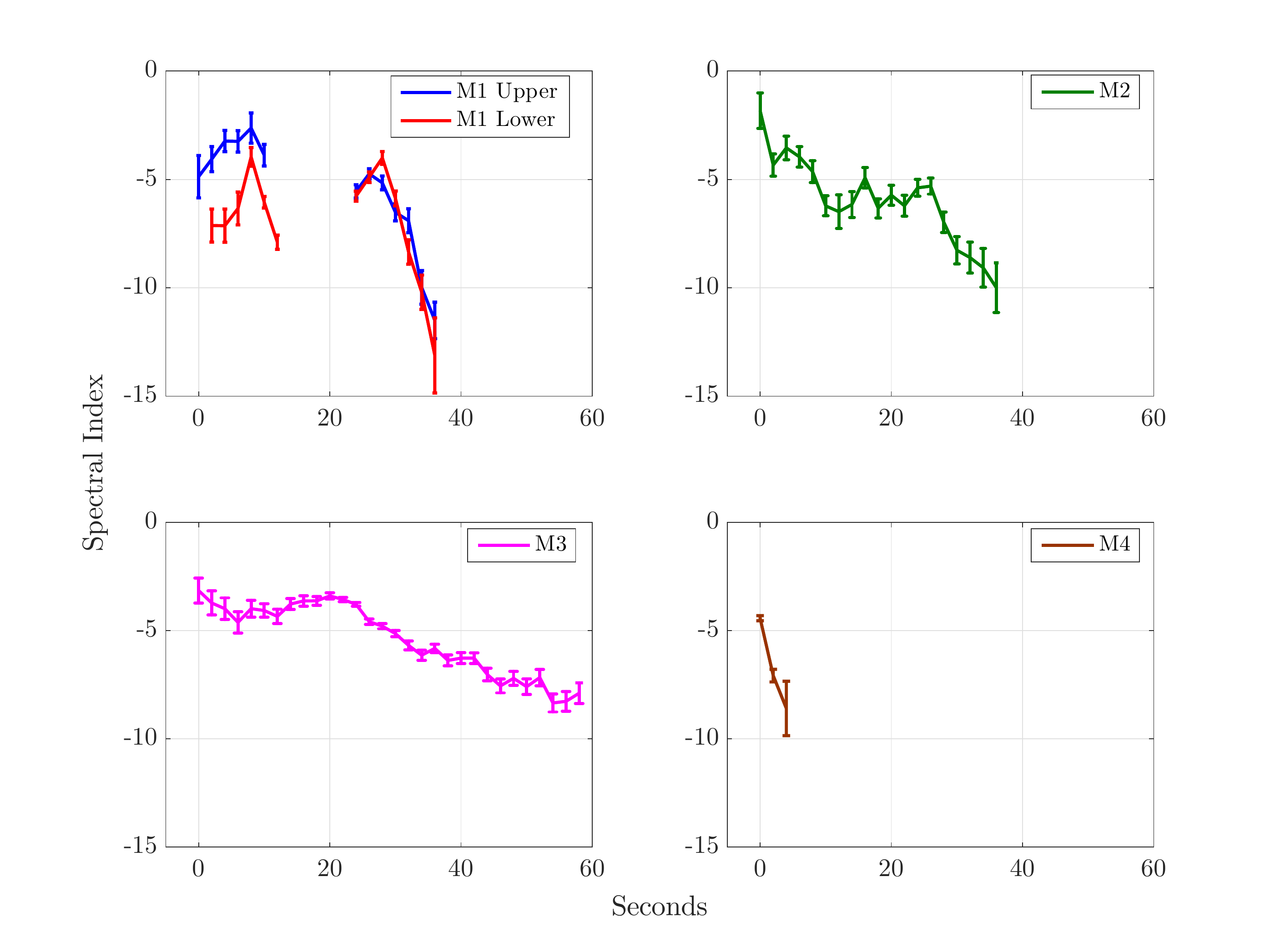}
	\caption{Spectral index as a function of time for each of the four meteor afterglows. M1 has two measurements, from the upper and lower part of the trail. The error bars represent the 95\% confidence level of the fits. }
	\label{fig:sindex_50}
\end{figure}

Figure \ref{fig:sindex_50} shows the spectral indices as a function of time for each meteor afterglow, and in the case of M1 Figure \ref{fig:sindex_50} shows indices from the upper and lower end of the trail. For each afterglow there is an apparent decrease (steepening) in index over time.

This trend is not surprising since the trail is diffusing outward and the electron number density is decreasing, therefore driving down the average plasma frequency, which is proportional to the square root of the electron number density. If the emission is coming from Langmuir waves, then this steepening of the spectra may be tracing the density profile evolution of the trail.

\subsection{Polarization of M4}

As mentioned in section \ref{section:ImageAnalysis}, the PASI images of M4 contained a significant amount of linear polarization. \citet{Obenberger15a} showed that the amount of instrumental polarization leakage in the PASI images was less than 10\% near zenith, for all Stokes parameters, therefore most of the observed polarization is intrinsic to M4.

\begin{figure}
	\centering
	\includegraphics[width = 5.5in]{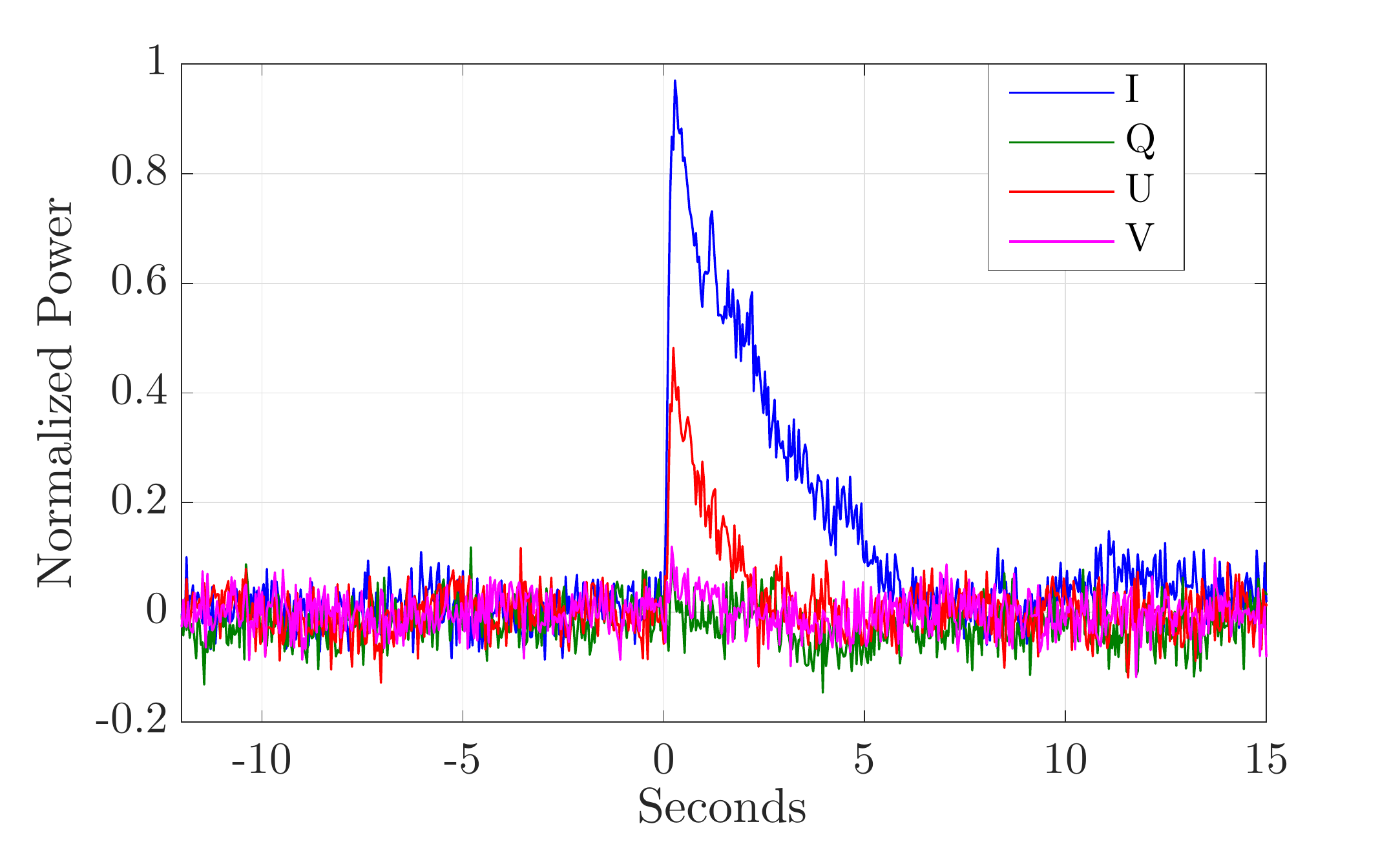}
	\caption{Normalized flux from M4, shown in Stokes I, Q, U and V. The emission reaches maximum brightness in 0.3 seconds and is observable for 5.8 seconds in Stokes I, and 2.4 seconds in Stokes U. There is no detectable emission in Stokes Q or V.}
	\label{fig:FB_4_I_U}
\end{figure}

The occurrence of M4 was serendipitously coincident with the day we modified the beam observations to record full Stokes parameters. We were able to confirm that the broadband emission of M4 was indeed polarized in Stokes U at around 50\% across all observed frequencies, and that there was no detectable stokes Q or V. However, since the LWA1 is not polarization calibrated, we cannot be precise with polarization percentage of M4. Figure \ref{fig:FB_4_I_U} shows the light curve of M4 in all Stokes polarizations.

It is certainly interesting that the radio afterglow of M4 contained linear polarization. To date, M1 was the only other event that has contained significant linear polarization. The polarized emission from M1 lasted for a relatively short duration, swept through frequency, and was not smooth across the observed frequencies. This clearly separated the polarized emission of M1 from the bulk the emission seen from any other meteor afterglow. 

The polarized emission from M4, on the other hand, is smooth across frequency and time and has a comparable spectral index to the other meteor afterglows. In all respects (aside from polarization) it is very similar to the unpolarized emission from M1, M2, and M3. Furthermore, all other meteor afterglows seen by PASI contain no detectable polarization. It should be noted that the orientation of the trail as measured by PASI is offset from north by $\sim 50^{\circ}$ to the east, this is near to the polarization angle of positive Stokes U ($45^{\circ}$ to the east).

\section{Observability at Higher Frequencies}

\subsection{Flux densities}

\begin{figure}
	\centering
	\includegraphics[width = 5.5in]{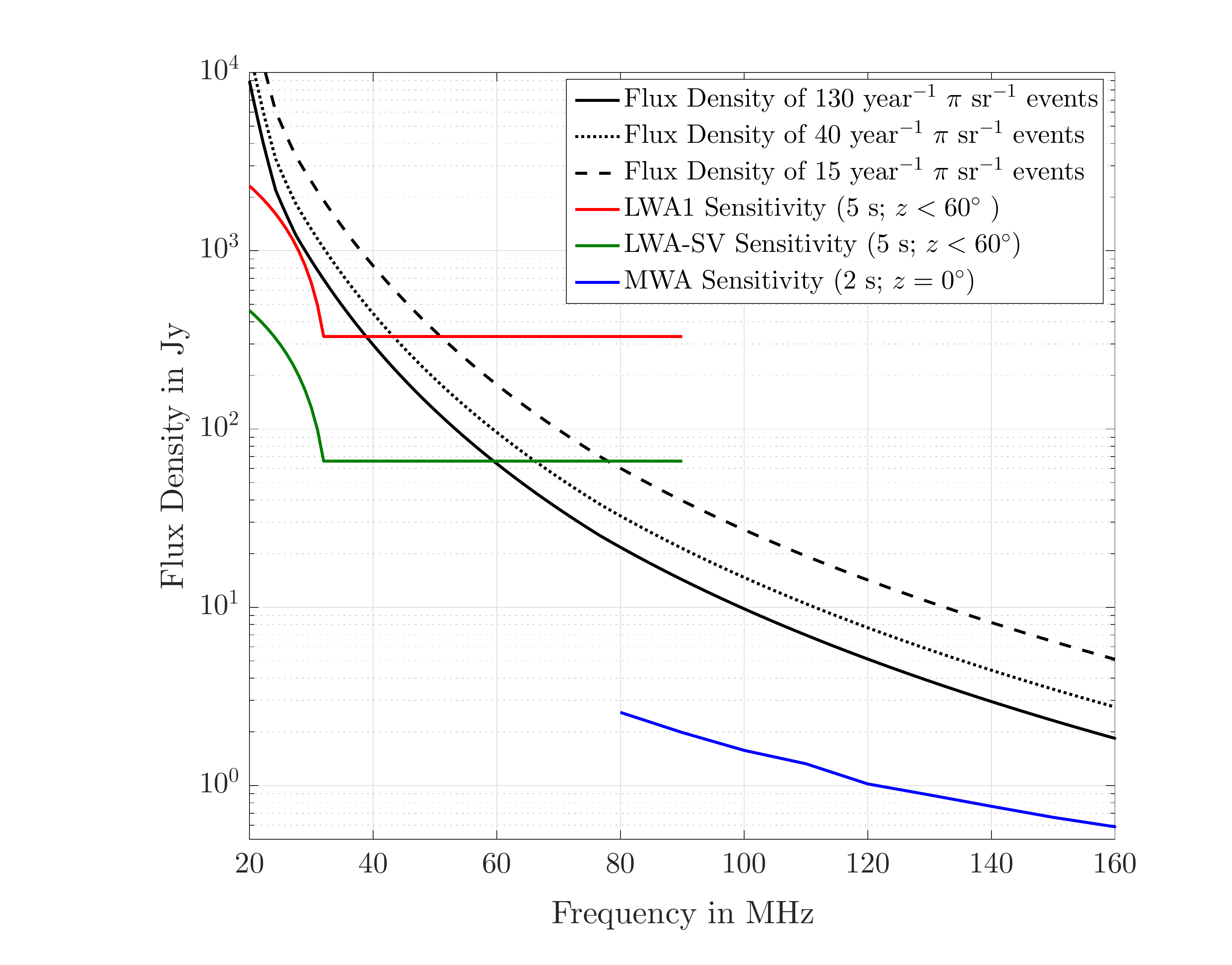}
	\caption{Extrapolated peak flux densities for M3, as well as events with minimum flux densities for events occurring 40 and 15 times year$^{-1}$ $\pi$ sr$^{-1}$. The peak flux density at each frequency does not follow a perfect power law, because the lower frequencies, peak at later times, where the spectra is steeper. Also shown are comparison plots of the $\sim 5 \sigma$ sensitivities of LWA1/PASI (5 second integration and 75 kHz bandwidth) and LWA-SV (5 second integration 2 MHz bandwidth) images, for zenith angles above 60 degrees, from 20 to 90 MHz \citep{Schinzel14,Obenberger15a} and the 5 $\sigma$ sensitivity of MWA images (2 second integrations and 30 MHz bandwidth) at zenith from 80 to 160 MHz \citep{Sutinjo15}.}
	\label{fig:VHF_extrap}
\end{figure}

To give an idea for what flux densities to expect at higher frequencies we have used the spectral index measurements from M3 to extrapolate the peak flux densities for typical meteor afterglows from 20 to 160 MHz. Assuming the rate measurements from section \ref{section:ImageAnalysis}, Figure \ref{fig:VHF_extrap} shows the flux densities, above which events should roughly occur 130, 40, and 15 times per year$^{-1}$ $\pi$ sr $^{-1}$. These curves give an idea of what to expect at higher frequencies, and provide an estimate of the required sensitivity, field of view, and observing time needed to detect the emission. 

\subsection{Potential for Other Observatories}

Along with the peak flux density Figure \ref{fig:VHF_extrap} also shows an estimated 5 $\sigma$ sensitivities, at a zenith angle of 60$^{\circ}$, of both PASI images (5 second integrations and 75 kHz bandwidth) and LWA-SV images (5 second integrations and 2 MHz bandwidth), and at zenith for MWA image images (2 second integrations and 30 MHz bandwidth). The LWA1 values are estimated from measurements above 32 MHz from \citet{Schinzel14} and \citet{Obenberger15a} and the measured value from 25.6 MHz done for this paper using PASI images. The sensitivity of the LWA1 is more or less constant above 32 MHz, but drops off below 32 MHz due to the response of the antennas over frequency. LWA-SV values are extrapolated from LWA1 measurements assuming a 2 MHz bandwidth. The MWA values were calculated using requested data from Adrien Sutinjo \citep{Sutinjo15}. With a image noise of 10 Jy at zenith the sensitivity for AARTFAAC is comparable to PASI over most of its frequency range \citep{Prasad14,Obenberger15a}, and being very similar in design, LWA-OVRO is comparable to LWA-SV. 

It is interesting to note that we have operated PASI for well over 1000 hours at both 52 and 74 MHz with no detections of meteors. This is consistent with the fact that the image noise at these frequencies is greater than the extrapolated flux densities for even the brightest (15 per year) events. The non-detection of meteors at 52 and 74 MHz can be used as an upper limit and are consistent with the power law fit. 

While PASI and AARTFAAC do not have the sensitivity to measure spectra above 50 MHz, LWA-SV and LWA-OVRO will be able to make broadband all-sky images which will surpass PASI's sensitivity. These facilities should be able to make broadband measurements of many meteors and extend the work presented in this paper to characterize the spectra. 

With its exceptionally high sensitivity, the MWA has the best opportunity to test the high frequency predictions of meteor radio afterglows made in this paper. As Figure \ref{fig:VHF_extrap} shows, the MWA should have high signal to noise from 80 to 160 MHz, but these predicted flux densities for the MWA frequency range assume that the emission will not be resolved. 

This may not be the case since the MWA has $\sim$ 4.3 arc minute resolution at 80 MHz, which is nearly 60 times finer than the LWA1 resolution at 38 MHz ($\sim 4.5^{\circ}$). If the emission is coming from a spherical region with a diameter that is just under the resolution limit of the LWA1 ($\sim$ 8 km at 100 km range), then the flux density as observed by the MWA would be 3600 times dimmer than that shown in Figure \ref{fig:VHF_extrap}. However, the 4.3 arc minute resolution of the MWA translates to roughly 125 meters at 100 km distance, which is about the altitude of a typical radio emitting meteor. Since it is unlikely that in the first 30 seconds the width of the meteor trail would extend more than about $\sim 100$ meters \citep{Ceplecha98,Jones91,Dyrud01}, the observed flux density could only be spread over the length of the trail. This would imply that the flux density could only be smaller by a factor of 60 assuming that the radiation is uniformly coming from an extended region 8 km long. A drop in flux density by a factor of 60 would place the emission of even the exceptionally bright events just out of reach for the MWA. 

While \citet{Obenberger14} clearly showed that in some cases the radiation can come from extended regions of the trail, the detailed structure is unknown. It could certainly be that the emission is being radiated from many small, isolated regions along the trail, rather than uniformly from the trail. If the case is the former then the MWA will stand a better chance of detecting the radio afterglows. 

\subsection{Near-Field Complications}

Another point we should stress concerns issues related to near-field vs. far-field interferometry. Most radio telescopes focus at infinity, because the vast majority of objects of interest are in the far-field, or effectively at an infinite distance. Focusing to infinity means that the correlator assumes that the incoming waves are all effectively plane waves. However, when objects are relatively close to the telescope this assumption does not hold, and special delays need to be introduced for each array element to account for the fact that the incoming waves are spherical rather than planar. It is standard procedure for all instruments mentioned in this paper to focus to infinity when producing their cross correlation products (visibilities).

The accepted transition from near-field to far-field occurs at the Fraunhofer distance, that is when the distance to an object is equal to the Fraunhofer area of a telescope $d = 2D^{2}/\lambda$, where $d$ is the distance to the object, $D$ is the diameter of the telescope, and $\lambda$ is the wavelength. For baselines separated by the Fraunhofer distance, the phase error is exactly equal to 90$^{\circ}$, meaning that the voltages completely cancel out. Any object that is closer than this will have some coherence. For the LWA1 and LWA-SV at 50 MHz this distance is $\sim$ 3 km, and $\sim$ 12 km for AARTFAAC and the inner 200 meter core of LWA-OVRO at 50 MHz. These values are on to two orders of magnitude smaller than the distance to a typical meteor.

LWA-OVRO also has an extended network of outlier antennas, which increases the effective size of the array to $\sim$ 3 km. For these long baselines, objects closer than 3000 km are considered in the near field. This distance is an order of magnitude greater than the typical range to a meteor. Similarly for the 3 km maximum baselines of the MWA at 90 MHz, the near-field/far-field transition occurs at 5400 km.

 In order to properly observe a meteor afterglow with the full arrays, the MWA and LWA-OVRO would need to calculate the appropriate delays, based on the position of the meteor, before correlation. This implies that they would need to know the position of the meteor before they correlate the raw voltages. It should be noted that they could apply the phase errors after correlation, however this would only fix the focus. They would still lose sensitivity due to decoherence of the correlated incoming waves.

Since both arrays contain a dense inner core, it is possible that they could identify the location of a meteor radio afterglow using a subset (or even several subsets) of antenna elements. Provided the raw voltages are stored, the data could be reprocessed later with all of the baselines to achieve higher resolution at full sensitivity.

As mentioned above the inner core of LWA-OVRO would be perfectly capable of observing meteors in the far-field. However, in order for the MWA (at 90 MHz) to observe a meteor at a distance of $\sim$ 100 km in the far-field, the maximum baseline for a subset of elements would need to be $\leq$ 400 m. Removing baselines above 400 m would decrease the sensitivity of the images, but would provide a simple means to use the standard correlator output.

\section{Conclusions}

We have reported yearly averaged meteor radio afterglows rate densities of 130, 40, and 15 events per year$^{-1}$ $\pi$ sr $^{-1}$, for events brighter than 1,700 Jy at 25.6 MHz and 540 and 1,000 Jy at 38.0 MHz. These rates are based on 1,135 hours of observing at 25.6 MHz and 12,876 hours at 38.0 MHz.  

We have also fit the dynamic spectra of four events to time dependent power laws. The spectral indices for all four events get steeper with time. These observations are qualitatively consistent for a diffusing plasma, assuming that the radiation is coming from Langmuir waves, where higher plasma frequencies would decay to lower plasma frequencies as the electrons diffuse into the ambient atmosphere. A comparison of these observations to a numerical model is warranted. 

Using the spectral indices of power law fits, we extrapolated higher frequency spectra for typical meteors occurring at the rates mentioned above. We then compared the expected flux densities to the sensitivities of other radio telescopes. In particular the MWA telescope has a unique opportunity to study the high frequency extension of meteor radio afterglow spectra, although high angular resolution and near-field effects will no doubt pose a challenge.

We have also reported on linear polarization detected in M4, steady polarization across all observed frequencies has not been observed in any other meteor radio afterglow. The majority of the observed polarization ($\sim$ 50\%) is mot likely intrinsic to the afterglow itself, rather than caused by instrumental leakage. M4 was distinct from most meteors, being dim (480 Jy at 34 MHz) and short in duration ($\sim$ 6 seconds). By observing at 34 MHz PASI is optimized to see dim events. Further observations will prove useful for polarization studies of meteor radio afterglows.

\begin{acknowledgments}
This research was performed while the author, Kenneth Obenberger, held an NRC Research Associateship award at the Air Force Research Laboratory, Kirtland AM, NM

Construction of the LWA1 has been supported by the Office of Naval Research under Contract N00014-07-C-0147. Support for operations and continuing development of the LWA1 is provided by the National Science Foundation under grants AST-1139963 and AST-1139974 of the University Radio Observatory program. 

All of the LWA1 data used in this article is publicly available at the LWA1 Data Archive (lda10g.alliance.unm.edu).

\end{acknowledgments}

\end{article}
%
%
%
%
%
%
%
%


\end{document}